\documentclass[aps,twocolumn,nofootnoteinbib,showPACS]{revtex4}
\usepackage{amsfonts}
\usepackage{amsmath}
\usepackage{amssymb}
\usepackage{graphicx}
\begin{document}
\preprint{ }
\title{Negativity and contextuality are equivalent notions of nonclassicality}
\author{Robert W. Spekkens}
\affiliation{Department of Applied Mathematics and Theoretical
Physics, University of Cambridge, Cambridge, United Kingdom CB3 0WA}


\begin{abstract}
Two notions of nonclassicality that have been investigated
intensively are: (i) negativity, that is, the need to posit negative
values when representing quantum states by quasiprobability
distributions such as the Wigner representation, and (ii)
contextuality, that is, the impossibility of a noncontextual hidden
variable model of quantum theory (also known as the
Bell-Kochen-Specker theorem). Although both of these notions were
meant to characterize the conditions under which a classical
explanation cannot be provided, we demonstrate that they prove
inadequate to the task and we argue for a particular way of
generalizing and revising them. With the refined version of each in
hand, it becomes apparent that they are in fact one and the same. We
also demonstrate the impossibility of noncontextuality or
nonnegativity in quantum theory with a novel proof that is symmetric
in its treatment of measurements and preparations.

\end{abstract}
\maketitle

It is common to assert that the discovery of quantum theory
overthrew our classical conception of nature. But what, precisely,
was overthrown? Being specific about the way in which a quantum
universe differs from a classical universe is a notoriously
difficult task and continues to be a subject of ongoing research
today. This problem has become one of practical concern in quantum
information theory~\cite{Nie00} and quantum metrology~\cite{Gio04}
as insights into the differences between the two theories help to
identify and analyze information-processing tasks for which quantum
protocols outperform their classical counterparts. The two notions
of nonclassicality with which we shall be concerned in this article
are negativity and contextuality, or more precisely, the presence of
negative values in quasi-probability representations of quantum
theory~\cite{Lut95} and the impossibility of noncontextual hidden
variable models~\cite{BKS}. We argue that these notions, construed
in the traditional manner, are not sufficiently general and we
promote a particular way of generalizing and revising them. In
particular, we argue that nonnegativity in the distributions
representing quantum states is not sufficient for classicality; the
conditional probabilities representing \emph{measurements} must also
be nonnegative. Furthermore, we argue that a classical explanation
cannot be ruled out by considering a single quasiprobability
representation, such as the Wigner representation~\cite{Wig32};
negativity must be demonstrated to hold for \emph{all} such
representations. Following previous work by the
author~\cite{Spek05}, we also argue that an assumption of
determinism that is part of the traditional notion of
noncontextuality should be excised and that context-independence
should be required not just for measurement procedures but for
preparation procedures as well. Under these refinements, the two
notions of nonclassicality are revealed to be equivalent.

\textbf{Negativity.} In 1932, Wigner showed that one can represent a
quantum state by a function on phase space, now known as the Wigner
function, having the property that the marginals over all
quadratures (linear combinations of position and momentum) reproduce
the statistics for the associated quantum observables~\cite{Wig32}.
This function cannot, however, be interpreted as a probability
distribution over a classical phase space because for some quantum
states it is not everywhere nonnegative.  We shall say that such
quantum states exhibit \emph{negativity} in their Wigner
representation. It is commonly thought that such negativity is a
good notion of nonclassicality. However, we argue that it is neither
a necessary nor a sufficient condition for the failure of a
classical explanation.

First, we show that it is not a necessary condition. It is
well-known that the original Einstein-Podolsky-Rosen two-particle
state has a positive Wigner representation \cite{BellonWig} so that
it can be associated with a classical probability density over the
phase space of the two particles (i.e. over local hidden variables).
However, it has also been shown that it is possible to violate a
Bell inequality with such a state \cite{BanWod}. How can this be?
The resolution of the puzzle is that one can only have a classical
interpretation of an experiment if both the preparations \emph{and
the measurements} admit a classical interpretation, and in the
experiments in question, the measurements that one requires -- such
as parity measurements -- do not admit such an interpretation
because the Wigner representations of the projectors have values
outside of the interval $[0,1]$ and consequently cannot be
interpreted as conditional probabilities. (A similar argument has
been made in Ref.~\cite{Rev05}.)

Neither is the negativity of the Wigner representation
\emph{sufficient} for nonclassicality. For example, if one considers
a limited set of preparations and measurements for which the
associated density operators and positive operator valued measures
(POVMs) are diagonal in some orthogonal basis, then the diagonal
components may be interpreted as classical probabilities, yielding a
classical explanation of the experimental statistics. Nonetheless,
if the diagonalizing basis does not consist of quadrature
eigenstates -- for instance, if it consists of number eigenstates --
then the Wigner representations of these preparations and
measurements will not be positive.  More generally, negativity of
the Wigner representation does not demonstrate that there isn't some
other representation with respect to which one achieves a classical
explanation. Note that what is classical about these explanations is
their use of probability theory. We allow the space of physical
states over which the probabilities are defined to be arbitrary.

\textbf{Generalizing the notion of negativity. } The lesson of the
above examples is that in evaluating the possibility of a classical
explanation of an experiment, one must consider the negativity of
not just the representation of preparations but of measurements as
well, and one must look at representations other than that of
Wigner.

There is a natural class of representations that includes the Wigner
representation and that allows one to preserve a notion of
nonclassicality as negativity. We call these ``quasiprobability
representations'' and define them by the following features. Every
density operator $\rho$, a positive trace-class operator on a
Hilbert space $\mathcal{H}$, is represented by a normalized and
real-valued function $\mu_{\rho}$ on a measurable space $\Lambda.$ \
That is, $\rho\leftrightarrow \mu_{\rho}(\lambda)$ where
$\mu_{\rho}:\Lambda\rightarrow\mathbb{R}$ and
$\int\mu_{\rho}(\lambda)d\lambda=1.$ \ Similarly, every POVM
$\{E_{k}\}$, a set of positive operators on $\mathcal{H}$ that sum
to identity, is represented by a set $\{\xi_{E_{k}}\}$ of
real-valued functions on $\Lambda$
that sum to the unit function on $\Lambda$. \ That is, $\{E_{k}%
\}\leftrightarrow\{\xi_{E_{k}}(\lambda)\}$ where
$\xi_{E_{k}}:\Lambda\rightarrow \mathbb{R}$ and
$\sum_{k}\xi_{E_{k}}(\lambda)=1$ for all $\lambda\in\Lambda.$ (The
trivial POVM $\{I\}$ is represented by $\xi_{I}(\lambda)=1$, and the
zero operator is represented by the zero function) Finally, the
representation must be such that
\begin{equation}
\mathrm{Tr}\left(  \rho E_{k}\right)  =\int\mathrm{d}\lambda\mu_{\rho}%
(\lambda)\xi_{E_{k}}(\lambda).\label{Bornforquasiprobrepn}%
\end{equation}
There are infinitely many such representations one could define, but
popular alternatives to Wigner include the Q and P representations
of quantum optics.

We define a \emph{nonnegative} quasiprobability representation of
quantum theory as one for which
\begin{equation}
\mu_{\rho}(\lambda) \geq0, \quad \xi_{E}(\lambda)
\geq0\label{positivityforPOVMs}
\end{equation}
for all density operators $\rho$ and all positive operators $E$ less
than identity (i.e. all possible POVM elements). This would
constitute a classical representation of all possible preparations
and measurements.

\textbf{Contextuality. }\strut The traditional notion of a
noncontextual hidden variable model of quantum theory can be
expressed as follows~\cite{Spek05}. Denoting a complete set of
variables in the model (what Bell refers to as the "beables" of the
model \cite{Belllocalbeables}) by $\lambda$, and the measurable
space of these by $\Lambda,$ one represents every pure quantum state
$|\psi\rangle$ by a normalized probability density on $\Lambda,$
$\mu_{\psi}(\lambda)$, and every projector-valued measure
$\{\Pi_{k}\}$ (the spectral
elements of a Hermitian operator) by a set $\{\xi_{\Pi_{k}%
}(\lambda)\}$ of $\{0,1\}$-valued indicator functions on $\Lambda$.
An indicator function $\xi_{\Pi_{k}}(\lambda)$ specifies the
probability of outcome $k$ given $\lambda$. Because \textit{some}
outcome must occur, indicator functions associated with a complete
set of outcomes sum to 1, i.e. $\sum_{k}\xi _{\Pi_{k}}(\lambda)=1$
for all $\lambda$. A \textit{$\{0,1\}$-valued} indicator function is
one for which $\xi_{\Pi_{k}}(\lambda) \in \{0,1\}$, so that the
outcome of the measurement is determined by $\lambda$ (rather than
being probabilistic).  We refer to this restriction on indicator
functions as the assumption of \emph{outcome determinism for sharp
measurements.} Note that this assumption is part of the traditional
notion of noncontextuality (a point to which we shall return).
Finally, in order for the hidden variable model to reproduce the
probability of outcome $k$ given $\psi$, one requires that
$\int\text{d}\lambda\,\mu_{\psi}(\lambda)\xi_{\Pi_{k}}(\lambda
)=\langle\psi|\Pi_{k}|\psi\rangle$.

To see why such a model is called \textquotedblleft
noncontextual\textquotedblright, note that whenever one of the
$\Pi_{k}$ has rank two or greater, it can be decomposed into a sum
of smaller rank projectors in many different ways, and each of these
corresponds to a different way of implementing the measurement -- a
different context. The representation of the measurement in the
hidden variable model is presumed to depend \emph{only} on the
$\Pi_{k}$, and not on how the measurement was implemented. The
representation is therefore independent of the context, hence
\emph{noncontextual}. The Bell-Kochen-Specker theorem establishes
that such a representation of quantum theory is impossible
\cite{BKS}.

\textbf{Generalizing the notion of contextuality.} As argued in
detail in previous work~\cite{Spek05}, the issue of whether a
measurement's representation in the model is context-dependent or
not can and should be separated from the issue of whether the
outcome of the measurement is determined uniquely or only
probabilistically by $\lambda$. Thus, whereas traditionally the
question of interest is whether or not the measurement outcome for a
given $\lambda$ depends on the context of the measurement, we claim
that the interesting question is whether the \emph{probabilities} of
different outcomes for a given $\lambda$ depend on the context. This
is analogous to Bell's generalization of the notion of locality from
measurement outcomes being causally independent of parameter
settings at space-like separation to the \emph{probabilities} of
measurement outcomes being so~\cite{Belllocalbeables}.
Mathematically, the proposed generalization corresponds to dropping
the assumption of outcome determinism for sharp measurements from
the definition of noncontextuality (by not requiring that
$\xi_{\Pi_{k}}(\lambda)$ be $\{0,1\}$-valued).

The fully general definition of a noncontextual hidden variable
model requires that \emph{all} procedures -- preparations,
transformations, and both projective and non-projective measurements
-- are represented in a manner that depends only on how the
procedure is represented in the quantum formalism. If two procedures
differ in ways that are not reflected in the quantum formalism we
call this difference part of the \emph{context} of the procedure.

To be specific, the assumption of \emph{preparation
noncontextuality} is that the probability distribution
$\mu_{\textrm{P}}(\lambda)$ associated with a preparation procedure
\textrm{P} depends only on the density operator $\rho$ associated
with \textrm{P}, i.e.
$\mu_{\mathrm{P}}(\lambda)=\mu_{\rho}(\lambda).$ \ For instance, if
the preparation is a mixture of pure states $|\psi_k\rangle$ with
weights $w_k$, then the distribution depends only on the average
density operator $\rho=\sum_k w_k |\psi_k\rangle \langle \psi_k|$
and not the particular ensemble. Similarly, the assumption of
\emph{measurement noncontextuality} is that the indicator function
$\xi_{\mathrm{M},k}(\lambda)$ representing outcome $k$ of a
measurement procedure $\text{M}$ depends only on the associated POVM
element $E_{k}$, i.e.,
$\xi_{\mathrm{M},k}(\lambda)=\xi_{E_{k}}(\lambda).$

In order to highlight the content of the assumption of
noncontextuality, it is useful to formalize the assumptions that
define a hidden variable model of quantum theory.  In fact, the set
of models that we characterize includes the case wherein the quantum
state is a complete description of reality and so it is better to
refer to these simply as ``ontological models''. Every preparation
procedure $P$ that is permitted by the theory is represented by a
normalized and positive function on a measurable space $\Lambda$,
and every measurement procedure $M$ is represented by a set of
positive functions on $\Lambda$ that sum to unity. Specifically, we
have $P\leftrightarrow\mu_{P}(\lambda),$ where
$\mu_{P}:\Lambda\rightarrow \mathbb{R}$ such that $\int_\Lambda
\mu_{P}(\lambda)d\lambda=1$, and
$M\leftrightarrow\{\xi_{M,k}(\lambda)\}$, where
$\xi_{M,k}:\Lambda\rightarrow\mathbb{R}$ such that $\sum
_{k}\xi_{M,k}(\lambda)=1$ for all $\lambda \in \Lambda$, and
\begin{equation}
\mu_{P}(\lambda)\geq0,\quad
\xi_{M,k}(\lambda)\geq0\,.\label{normforpreps}
\end{equation}
Finally, let $\mathcal{P}_{\rho}$ denote all preparation procedures consistent
with the density operator $\rho$, and $\mathcal{M}_{\{E_{k}\}}$ all
measurement procedures consistent with the POVM $\{E_{k}\}.$ An ontological
model of quantum theory is such that for all $P\in\mathcal{P}_{\rho},$\textbf{
}and for all $M\in\mathcal{M}_{\{E_{k}\}},$
\begin{equation}
\int d\lambda\mu_{P}(\lambda)\xi_{M,k}(\lambda)=\mathrm{Tr}\left(
\rho E_{k}\right)  .\text{ }\label{Bornruleforontologicalmodel}
\end{equation}

A \emph{noncontextual} ontological model of quantum theory (in our
generalized sense) is an ontological model that satisfies
\begin{align}
\mu_{P}(\lambda) &  =\mu_{\rho}(\lambda)\text{ for all }P\in\mathcal{P}_{\rho
}\label{PNC}\\
\xi_{M,k}(\lambda) &  =\xi_{E_{k}}(\lambda)\text{ for all }M\in\mathcal{M}%
_{\{E_{k}\}}.\label{MNC}%
\end{align}
Eqs.~(\ref{PNC}) and (\ref{MNC}) codify the assumptions of
noncontextuality for preparations and measurements respectively.

Whatever reasons one can provide in favor of the assumption of
measurement noncontextuality (for instance, that it is the simplest
possible explanation of the context-independence of the right-hand
side of Eq.~(\ref{Bornruleforontologicalmodel})), the very same
reasons can be given in favor of the assumption of preparation
noncontextuality. Thus if one takes noncontextuality for
measurements as a condition for classicality, then noncontextuality
for preparations should also be required.

\textbf{Equivalence of the two notions of nonclassicality.} By
substituting the conditions for preparation and measurement
noncontextuality, Eqs.~(\ref{PNC}) and (\ref{MNC}), into the
conditions for an ontological model, Eqs.~(\ref{normforpreps}) and
(\ref{Bornruleforontologicalmodel}), we obtain the conditions for a
nonnegative quasi-probability representation of quantum theory,
Eqs.~(\ref{Bornforquasiprobrepn}) and (\ref{positivityforPOVMs}). So
we see that by these definitions, a noncontextual ontological model
of quantum theory exists if and only if a nonnegative
quasi-probability representation of quantum theory exists.

What we have discovered by this analysis is that the assumption of
noncontextuality (in our generalized sense) has always been implicit
in the notion of a quasi-probability representation. Given its
conceptual significance and mathematical simplicity, it is
surprising that this connection has not been noted previously. Two
likely reasons for this are: (i) the lack of emphasis on the
representation of measurements in discussions of negativity, and
(ii) the lack of a generalization of contextuality to preparations
and nonprojective measurements and the failure to distinguish the
assumption of measurement noncontextuality from that of outcome
determinism.

\textbf{No-go theorems for nonnegativity or noncontextuality.} An
ontological model of quantum theory that is noncontextual, in the
generalized sense described here, is impossible~\cite{Spek05}. It
follows that a nonnegative quasiprobability representation of
quantum theory is also impossible. This fact is unlikely to surprise
those who know quantum theory well. Nonetheless, to our knowledge,
it has not been demonstrated previously (although
Montina~\cite{Mon06} came close to doing so, as we discuss in the
conclusions).

An unfortunate feature of existing no-go theorems for noncontextual
models is that they do not proceed \emph{directly} from the
assumption of generalized noncontextuality to a contradiction. For
instance, in Ref.~\cite{Spek05}, it is shown that one can base such
a proof on the Bell-Kochen-Specker theorem; however, the
contradiction is derived not only from the assumption of
noncontextuality for sharp measurements, but also the assumption of
outcome determinism for sharp measurements, and the latter
assumption is in turn derived from noncontextuality for
preparations.  So, despite the standard impression that these no-go
theorems concern only the representation of measurements, we see
that the representation of preparations enters the analysis in an
indirect way.
Similarly, in no-go theorems that appeal
only to the assumption of noncontextuality for preparation
procedures (\cite{Spek05}, Sec.~IV), one still relies on the fact
that only preparations associated with probability distributions
that are nonoverlapping can be discriminated by a single-shot
measurement. Thus the representation of measurements has appeared,
in an indirect way, within a proof based primarily on the
representation of preparations. A proof that is even-handed in its
treatment of preparations and measurements would be preferable and
we now provide one.

\textbf{An even-handed no-go theorem for nonnegativity or
noncontextuality. } Suppose that one has a set of preparation
procedures associated with density operators $\rho_{j}$. A procedure
associated with the mixture $\rho\equiv \sum_{j}w_{j}\rho_{j}$ can
be implemented as follows. Sample an integer $j$ from the
probability distribution $w_{j}$ (by rolling a die for instance),
and implement the preparation procedure associated with $\rho_{j}$.
If the distributions over $\lambda$ that represent each of these
procedures in the ontological model are denoted by
$\mu_{\rho_{j}}(\lambda)$, then clearly the distribution that
represents the mixture $\rho$ is $\mu_{\rho}(\lambda
)=\sum_{j}w_{j}\mu_{\rho_{j}}(\lambda)$.
Thus in a
noncontextual ontological model,
\begin{equation}
\text{if }\rho=\sum_{j}w_{j}\rho_{j},\text{ then
}\mu_{\rho}(\lambda)=\sum _{j}w_{j}\mu_{\rho_{j}}(\lambda),
\label{convexlinearityforstates}
\end{equation}
A similar argument concerning a mixture of measurements, each of
which has a distinguished outcome associated with a positive
operator $E_j$, establishes that in a noncontextual ontological
model,
\begin{equation}
\text{if }E=\sum_{j}w_{j}E_{j}\text{ then }\xi_{E}(\lambda)=\sum_{j}w_{j}%
\xi_{E_{j}}(\lambda). \label{convexlinearityforeffects}
\end{equation}

A real-valued function $f$ on the space $\mathcal{L}(\mathcal{H})$
of linear operators on $\mathcal{H}$ is \emph{convex-linear} on a
convex set $\mathcal{S}\subset\mathcal{L}(\mathcal{H})$ if
$f(\sum_{k}w_{k}R_{k})=\sum_{k} w_{k}f(R_{k})$ for $R_{k}\in
\mathcal{S}$ and $w_{k}$ a probability distribution over $k$.
Eqs.~(\ref{convexlinearityforstates}) and
(\ref{convexlinearityforeffects}) assert that $\mu$ as a function of
$\rho$ is convex-linear on the convex set of density operators, and
$\xi$ as a function of $E$ is convex-linear on the convex set of
positive operators less than identity (the set of ``effects'').

The first two steps of the no-go theorem are familiar as key
elements of the generalization of Gleason's theorem~\cite{Gle57} to
POVMs~\cite{Bus03,CFMR04}
and analogous reasoning plays an important role in Hardy's
axiomatization of quantum theory \cite{Har01}. The first step is to
note that a function $f$ that is convex-linear on a convex set
$\mathcal{S}$ of operators that span the space of Hermitian
operators (and that takes value zero on the zero operator if the
latter is in $\mathcal{S}$) can be uniquely extended to a linear
function on this space. Specifically, if $A$ is a Hermitian operator
that can be decomposed as $A=\sum_{k}a_{k}R_{k}$, where $R_k \in
\mathcal{S}$, then the extension is $f(A)=\sum_{k}a_{k}f(R_{k})$
\footnote{To see that it is unique, note that if
$A=\sum_{j}a'_{j}R'_{j}$ is a distinct decomposition of $A$, then
$\sum_{k|a_k\ge0}a_{k}R_{k}+\sum_{j|a'_j<0}|a_{j}|R'_{j} =
\sum_{k|a_k<0}|a_{k}|R_{k}+\sum_{j|a'_j\ge0}a_{j}R'_{j}$. Dividing
this equation by $C=\sum_{k|a_k\ge0}a_{k}+\sum_{j|a'_j<0}|a_{j}|$,
both sides become convex-linear functions of trace-class positive
operators from which is follows that
$\sum_{k}a_{k}f(R_{k})=\sum_{j}a'_{j}f(R'_{j})$.}.

The second step is to note that by Reisz's representation theorem,
the linear function $f(A)$ can be written as the Hilbert-Schmidt
inner product of $A$ with some fixed Hermitian operator, say
$B^{\dag},$ so that $f(A)=\mathrm{Tr}(AB).$

It follows that
\begin{equation}
\mu_{\rho}(\lambda)=\mathrm{Tr}[\rho F(\lambda)], \quad
 \xi_{E}(\lambda)=\mathrm{Tr}[\sigma(\lambda)E], \label{derivedlinearity1}
\end{equation}
where $F$ and $\sigma$ are functions from $\Lambda$ to the Hermitian
operators on $\mathcal{H}$.

A noncontextual ontological model, or equivalently, a nonnegative
quasiprobability representation, is one for which
$\mu_{\rho}(\lambda)\ge 0$ and $\xi_{E}(\lambda)\ge 0$ for all
$\rho$ and $E$, which implies that the operators $F(\lambda)$ and
$\sigma(\lambda)$ are not merely Hermitian but positive as well.
Given that $\int\mu_{\rho}(\lambda)\mathrm{d}\lambda=1$, it follows
that $\int F(\lambda)\mathrm{d}\lambda=I,$ and so we can conclude
that $F(\lambda)d\lambda$ is a POVM.
Furthermore, given that $\xi_I(\lambda)=1$ for all $\lambda$, we
also have \textrm{Tr}$[\sigma(\lambda)]=1$ and so we can conclude
that $\sigma$ is a map from $\Lambda$ to density operators.

We now show, using a proof by contradiction, that
$\mu_{\rho}(\lambda)$ and $\xi_E(\lambda)$ of this sort cannot
reproduce the quantum predictions.  To do so, $\mu_{\rho}(\lambda)$
and $\xi_E(\lambda)$ would need to satisfy
Eq.~(\ref{Bornforquasiprobrepn}), which implies, via
Eq.~(\ref{derivedlinearity1}), that $\int d\lambda \mathrm{Tr}[\rho
F(\lambda)]\mathrm{Tr}[\sigma(\lambda)E]=\mathrm{Tr}(\rho E)$ for
all $\rho$ and $E.$
There are many ways of deriving a contradiction
from here.
 Given that the set of density operators spans the space
$\mathcal{L}(\mathcal{H})$, we can infer that $E=\int d\lambda
\xi_{E}(\lambda) F(\lambda)$, i.e. every $E$ is a positive
combination of the $F(\lambda)$.  Now consider a POVM $\{ E_k \}$
with rank-1 elements.  Any given element $E_k$ is a positive
combination of the $F(\lambda)$, specifically, $E_k=\int d\lambda
\xi_{E_k}(\lambda) F(\lambda)$.  However, a rank-1 positive operator
admits only trivial positive decompositions into positive operators
(namely, into ones that are proportional to itself).  It follows
that $F(\lambda) \propto E_k$ for all $\lambda$ in the support of
$\xi_{E_k}$. Recalling that for every $\lambda \in \Lambda$, there
exists a $k$ such that $\lambda$ is in the support of $\xi_{E_k}$,
it follows that for every $\lambda \in \Lambda$, there exists a $k$
such that $F(\lambda) \propto E_k$. Repeating the argument for
another POVM with rank-1 elements, say $\{ E'_j\}$, we conclude that
for every $\lambda \in \Lambda$, there exists a $j$ such that
$F(\lambda) \propto E'_j$.  However, given that no element of $\{
E_k\}$ needs to be proportional to any element of $\{ E'_j\}$ (for
instance, they may be the projector-valued measures corresponding to
two bases having no elements in common), we arrive at a
contradiction. [Noting from Eqs.~(\ref{Bornforquasiprobrepn}) and
(\ref{derivedlinearity1}) that $\{F(\lambda)\}$ and
$\{\sigma(\lambda)\}$ are dual frames in the operator space, this
result implies that the dual of a frame of positive operators cannot
also be a frame of positive operators. A direct proof of this fact
is possible \cite{Fer07} and provides a faster route to the
contradiction.]

A similar argument to the one just provided can be found in the
recent work of Montina~\cite{Mon06}, where it is demonstrated that
to avoid negative probabilities in ontological models, the
representation of pure states cannot depend bilinearly on the
wavefunction.  Although the representation of mixed quantum states
is not discussed, it is a short step from this result to a
demonstration of a failure of preparation noncontextuality. The
reason is as follows.  As argued earlier, if $\mu_{\rho_i}(\lambda)$
is the representative of the procedure $P_i$ associated with the
pure state $\rho_i$, then the procedure that implements $P_i$ with
probability $w_i$ is represented by $\sum_i w_i
\mu_{\rho_i}(\lambda)$.  However, if, as Montina's work entails,
$\mu_{\rho_i}(\lambda)$ is not a linear function of $\rho_i$, then
$\sum_i w_i \mu_{\rho_i}(\lambda)$ cannot depend merely on the
density operator $\sum_i w_i \rho_i$ that is associated with the
mixed preparation, but must also depend on the context of this
preparation -- in particular, the pure state ensemble from which the
mixed state was formed. Therefore these results are seen to provide
another demonstration of the impossibility of a model that is
noncontextual in the generalized sense discussed here.

Attempts to characterize nonclassicality from either the perspective
of hidden variables or that of quasiprobability representations
drive one to the same conclusion: that the only way in which one can
salvage the possibility of an ontological model
is to deny the implicit starting point of these representations, the
assumption of noncontextuality.

The author gratefully acknowledges J.~Barrett, E.~Galv\~{a}o,
L.~Hardy, M.~Leifer, A.~Montina, T.~Rudolph and J.~Sipe for
discussions on these issues.

\end{document}